\documentclass[twocolumn,prx,aps,superscriptaddress]{revtex4-2}
\usepackage{amssymb,amsmath,amsfonts}

\usepackage{url}
\usepackage[latin9]{inputenc}
\usepackage{amstext}
\usepackage{braket}
\usepackage{enumitem}
\usepackage{graphicx,bm,palatino}
\usepackage[colorlinks=true,linkcolor=blue,urlcolor=blue,citecolor=blue,pdfusetitle]{hyperref}

\usepackage[sc]{mathpazo} 

\usepackage{hyperref,cleveref}

\usepackage[dvipsnames]{xcolor}
\usepackage[caption=false]{subfig}

\usepackage{times}
\usepackage{bbm}
\makeatother

\begin{document}
\title{Spectral Density Classification For Environment Spectroscopy}
\date{\today}

\author{J. Barr}
\email{jbarr24@qub.ac.uk}
\affiliation{Centre for Quantum Materials and Technologies, School of Mathematics and Physics, Queen's University Belfast, BT7 1NN, United Kingdom}
\author{G. Zicari}
\email{G.Zicari@qub.ac.uk}
\affiliation{Centre for Quantum Materials and Technologies, School of Mathematics and Physics, Queen's University Belfast, BT7 1NN, United Kingdom}
\author{A. Ferraro}
\affiliation{Centre for Quantum Materials and Technologies, School of Mathematics and Physics, Queen's University Belfast, BT7 1NN, United Kingdom}
\affiliation{Dipartimento di Fisica ``Aldo Pontremoli'', Universit\`{a} degli Studi di Milano, I-20133 Milano, Italy}
\author{M. Paternostro}
\affiliation{Centre for Quantum Materials and Technologies, School of Mathematics and Physics, Queen's University Belfast, BT7 1NN, United Kingdom}

\begin{abstract}
Spectral densities encode the relevant information characterising the system-environment interaction in an open-quantum system problem. Such information is key to determining the system's dynamics. In this work, we leverage the potential of machine learning techniques to reconstruct the features of the environment. Specifically, we show that the time evolution of a system observable can be used by an artificial neural network to infer the main features of the spectral density. In particular, for relevant examples of spin-boson models, we can classify with high accuracy the Ohmicity parameter of the environment as either Ohmic, sub-Ohmic or super-Ohmic,  thereby distinguishing between different forms of dissipation.
\end{abstract}

\maketitle

\section{Introduction}
\label{sec:intro}

Recent progress in the field of quantum technologies has advanced our capabilities to control quantum systems and exploit their non-classical properties. Yet, this task presents significant challenges. Quantum systems are inherently open, as they inevitably interact with their surrounding environment~\cite{Breuer-Petruccione1,Rivas2012}. They are thus susceptible to gain and losses, as well as to the genuine quantum phenomenon of decoherence~\cite{ZurekRev:2003,Braun:2001,Strunz:2003}, which disrupts the phase coherence of superposition states, posing a major obstacle in preserving quantum states~\cite{Unruh:1995,Chuang:1995}. If we are to effectively devise strategies for mitigating adverse environmental influences on a system, it is crucial to have a comprehensive understanding of the effects that need to be addressed when facing open quantum dynamics. This, in turn, requires a full characterisation of the mechanism governing the system-environment interaction. 

To address such challenge, here we tackle the problem of characterising environmental effects on an open quantum system harnessing recent advances in the field of Machine Learning (ML). The latter has opened up new data-driven approaches, which have shown their effectiveness in various applications in the field of quantum technologies~\cite{Marquardt:2023}. Among those, some are very close to the spirit of this work. ML-based methodologies have been applied to quantum tomography~\cite{Palmieri2020, Torlai2018, Banchi_2018}, quantum channel discrimination~\cite{PhysRevA.106.032409}, simulation of open quantum systems~\cite{doi:10.1126/science.aag23021, PhysRevLett.128.090501, BANDYOPADHYAY2018272,  PhysRevLett.122.160401}, as well as quantum control~\cite{GIANNELLI2022128054, Niu2019,Sgroi2021}. Refs. \cite{10.1063/5.0035498, 10.1063/5.0157639} reported the deployment of deep-learning methods to the inference of photon correlation functions and phonon blockade effects based on homodyne-detection schemes.

 We focus on the typical open quantum system scenario, where we are able to effectively describe and control the reduced system, as opposed to the infinitely many uncontrollable environmental degrees of freedom which are responsible for dissipation and decoherence.
In this setting, we focus on the interaction of a given system with an external environment in terms of the Spectral Density (SD), which, by encoding full information about the system-environment coupling, allows us to determine the two-time correlation function of the environment. Having full knowledge of this quantity
allows us to predict the temporal behaviour of an open quantum system without a full microscopic description of the environment. 

The SD for a given system-environment interaction, however, is rarely directly available and challenging to calculate from first principles. The form of a SD is at best phenomenologically inferred through empirical data gathered from experimental observations, and at worst {\it guessed} using \emph{ad hoc} assumptions, which might result in significant discrepancies between the predicted and actual dynamics of the system~\cite{PhysRevLett.118.100401}. In this work, we consider the case of a quantum system interacting with a bosonic thermal bath.  Depending on the nature of the system-bath interaction, the system dynamics can manifest as either pure dephasing or amplitude damping~\cite{nielsen_chuang_2010,Benenti:2018,Quanta77}. The particular choice of the SD in this setting is responsible for possible memory effects. On one hand, we can encounter a scenario where the information is monotonically flowing from the system to the bath, i.e. the usual scenario characterising quantum Markovian processes~\cite{Gardiner:2009,CHRUSCINSKI20221}. On the other hand, some functional forms of the SD are suitable to model a physical situation in which the system, dynamically interacting with its environment, can partially retrieve the information that was previously lost --- these processes are dubbed as non-Markovian instead~\cite{Rivas:20141,BreuerRev:2016}.

 Prior works have studied the use of ML for noise characterisation in open quantum systems. Various methods have been explored, such as studying the noise in qubit systems using two-pulse echo decay curves~\cite{PRXQuantum.2.010316}, and random pulse sequences that are applied to the qubit~\cite{ Youssry2020}. Additionally, other studies have focused on constructing the power spectral density for ensembles of carbon impurities around a nitrogen vacancy centre in diamond~\cite{martina2023deep}, and inferring the environment affecting superconducting qubits~\cite{papivc2022neural2}.
 
In this work, we show that an artificial Neural Network (NN) can be used to \emph{classify} the SD characterising the dynamics of a system,
 based on its features. Previous research has examined the classification of {\it aspects} of noise in open quantum systems. For instance, in Ref.~\cite{martina2023machine} ML techniques were used to discern between Markovian and non-Markovian noise.  More pertinent to the matter at hand, aspects of the problem of distinguishing between Ohmic, sub-Ohmic and super-Ohmic SDs have already been studied: in Ref.~\cite{garau2019machine2}, a scenario where a probe qubit is used to access a second inaccessible one is proposed to infer the Ohmicity class by using NNs and leveraging the special features of quantum synchronisation. In Ref.~\cite{palmieri:2021}, a different use of NNs was put forward as tomographic data at just two instants of time were used, rather than a time-series approach. In contrast, this work takes a simpler approach by utilising the time evolution of a system observable for classification without the need for a probe system or tomographically complete information. We focus on the case of a general Spin-Boson (SB) model to show that, even when the environment cannot be exactly traced out to infer the reduced dynamics of a system, a NN can classify the SD with high accuracy. Furthermore, we discuss the limitations imposed by the fluctuation of the parameters in the SD, the number of sampled points in the time signals, and measurement sampling noise. Our study emphasises the potential of ML techniques to characterise environments with arbitrary SDs.

The remainder of this paper is structured as follows: in \Cref{sec:methods} we provide an introduction to the general setting under consideration, as well as the ML approach utilised. Specifically, we examine an arbitrary system that is interacting with a bosonic environment and we give some background on the ML model used, namely, NNs. Next, in \Cref{sec:models} we detail the physical models that are considered. We investigate two SB models: in the first case, we are able to exactly derive the pure dephasing dynamics starting from the full system-bath unitary evolution; in the second case, we work in the weak coupling limit to approximately derive the reduced dynamics of the system. In both cases, the dynamics can feature non-Markovian effects, depending on the SD we select. In \Cref{sec:results} we discuss the architecture of the NNs, along with a detailed discussion of the results of training and testing for each model. Finally, we give our conclusive remarks and discuss our future outlook in \Cref{sec:conclusions}.

\section{General setting and methods}
\label{sec:methods}

\begin{figure}
\centering\includegraphics[width=0.7\columnwidth]{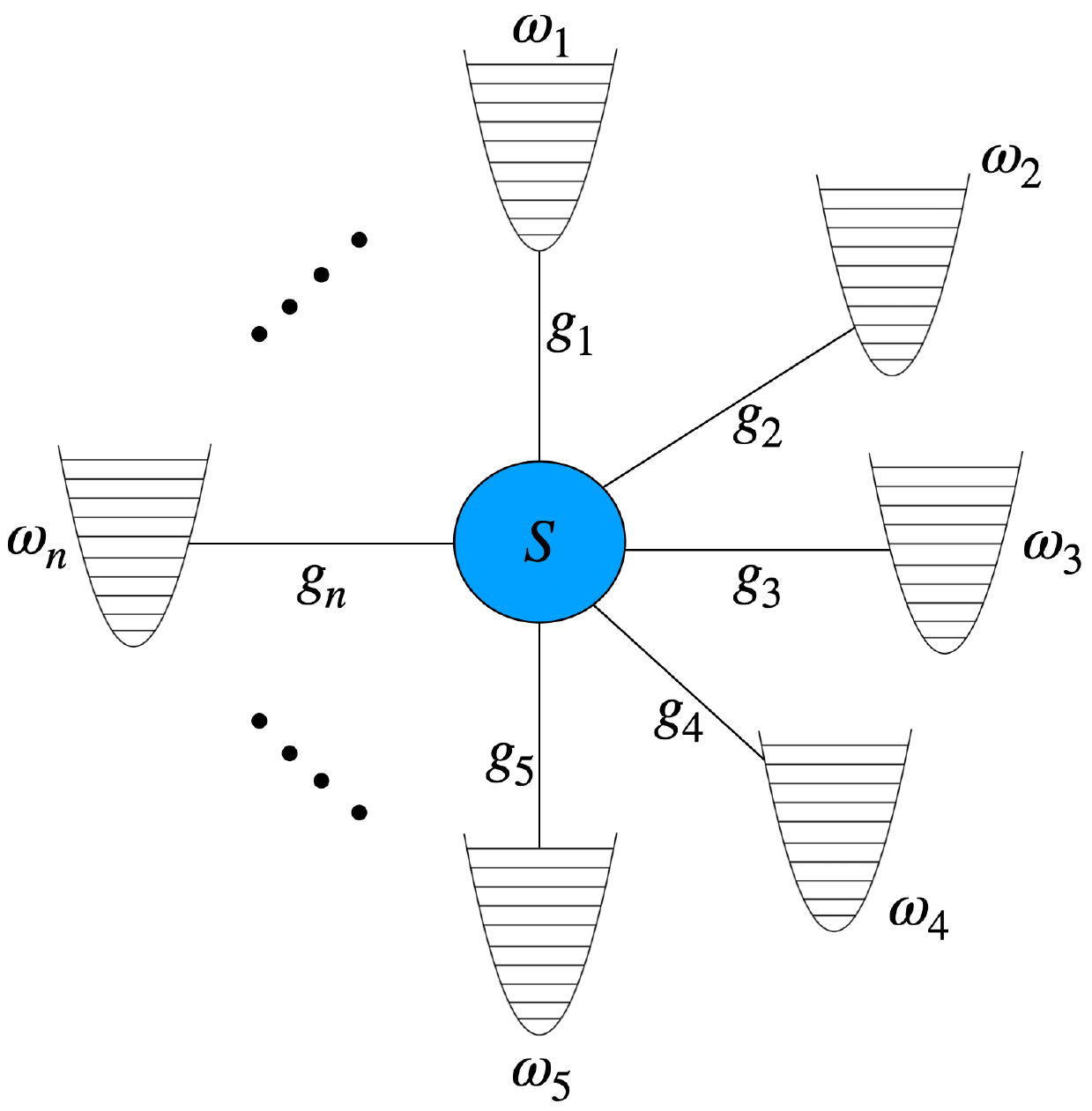}
	\caption{Sketch of a generic open quantum system $S$ interacting with a bosonic environment composed of infinitely many harmonic oscillators labelled by the integer $n$. Each oscillator has frequency $\omega_n$ and is coupled to the system at a rate $g_n$.}\label{fig:systemcoupledtobosonicenv}
\end{figure}

Let us consider the general setting of an arbitrary system interacting with an environment which is comprised of infinitely many bosonic modes, as shown in \Cref{fig:systemcoupledtobosonicenv}. This scenario reproduces the ubiquitous Caldeira-Leggett model, which describes the motion of a quantum particle undergoing a Brownian motion~\cite{Caldeira:1983,Leggett_rev:1987}. The full (time-independent) Hamiltonian reads as 

\begin{align}
\label{eq:Hamiltonian}
\hat{H} = \hat{H}_S + \hat{H}_B + \hat{H}_I,
\end{align}
where $\hat{H}_S$ and $\hat{H}_B$ are the Hamiltonian operators of the system and the environment, respectively. The system-environment interaction term $\hat{H}_I$ is expressed in the form
\begin{equation}
\label{eq:H_I}
    \hat{H}_I = \hat{X} \otimes \hat{B},
\end{equation}
where $\hat{X}$ is a generic system operator, while $\hat{B}$ is an operator of the bath. We take the latter as

\begin{equation}
\label{eq:bath_operator}
   \hat{B} = \sum_k \left( g_k \hat{b}_k^{\dagger} + g_k^* \hat{b}_k \right),
\end{equation}
where the coefficient $g_k$ accounts for the interaction strength between the $k$-th mode of frequency $\omega_k$, while $\hat{b}_k^{\dagger}$ and $\hat{b}_k$ are the creation and annihilation operators associated with it. The coupling coefficients enter in the formal definition of the SD, i.e. $J(\omega) = \sum_k |g_k|^2 \delta \left( \omega - \omega_k \right)$, the latter encoding all the information about the system-environment interaction. Since we are interested in the typical irreversible open system scenario, we will assume that the distribution of modes forms a continuum, so that the system dynamics does not display recurrences~\cite{Breuer-Petruccione1,Rivas:2010, Pucci:2013}. In this limit, the SD appears in the expression for the correlation function of a bosonic bath, defined as $\alpha_{\beta} (t) \equiv \langle \hat{B}(t) \hat{B}(0) \rangle_B$, where $\hat{B}(t)$ is the bath operator in the interaction picture with respect to the free Hamiltonian $\hat{H}_0 = \hat{H}_S + \hat{H}_B$. In Appendix \ref{app:A}, we show that if the environment is in a thermal Gibbs state, the correlation function can also be expressed as
\begin{equation}
\label{eq:bosonic_corr_f1}
\alpha_{\beta} (t) \equiv \nu (t) + i  \mu (t) \, ,
\end{equation}
where
\begin{align}
\label{eq:bosonic_corr_f2}
\begin{bmatrix} \nu(t) \\  \mu(t) \end{bmatrix} = \int\limits_{0}^{ \infty}  J(\omega) \begin{bmatrix}  \cos \left( \omega t \right) \coth \left( \frac{\beta \omega}{2} \right) \\ - \sin \left( \omega t \right) \end{bmatrix} \textrm{d} \omega \, ,
\end{align}
with $\beta = 1/T$. Note that hereafter we will work in units such that $\hbar=1$ and $k_B = 1$. The two functions $\nu(t)$ and $\mu(t)$ are also referred to as noise and dissipation kernels, respectively: the latter is independent of the temperature of the environment. The effective dynamics of the system, governed by a master equation, crucially depends on the correlation function $\alpha_\beta(t)$, which represents the fingerprint of the environment. The function $\alpha_\beta(t)$ is ultimately determined by the shape of the SD, which essentially contains all of the information about the environment needed to solve the dynamics of the system, and, thus, obtain the time evolution of any of its observables. The expectation value of a generic system observable at time $t$ is indeed given by
\begin{align}
\label{eq:observable}
\langle \hat{O}(t) \rangle \equiv \operatorname{Tr}_{SB} \left ( \hat{O} e^{- i \hat{H} t} \,  \hat{\rho}_{SB}^0 \, e^{i \hat{H} t}\right ),
\end{align}
where $\hat{H}$ is the system-environment Hamiltonian of \Cref{eq:Hamiltonian}, while the global initial state is factorised as $\hat{\rho}_{SB}^0 = \hat{\rho}^{0} \otimes \hat{\rho}_B$, with $\hat{\rho}^{0}$ and $\hat{\rho}_B$ being the initial system and environmental states, respectively. We assume the environment to be given by a large bosonic thermal reservoir, i.e. $\hat{\rho}_B = e^{- \beta \hat{H}_B}/ \mathcal{Z}_B$, where $\mathcal{Z}_B \equiv \operatorname{tr}_B (e^{- \beta \hat{H}_B})$ is the reservoir partition function. Under these hypotheses, it can be shown that the only environmental quantity entering in the expression of $\langle \hat{O}(t) \rangle$ is the SD $J(\omega)$. 

Here we focus on special classes of SDs, which can be expressed as~\cite{deVega_Alonso_rev:2017, Leggett_rev:1987}
\begin{equation}
\label{eq:Jexp}
    J(\omega) = \eta \omega_c^{1-s} \omega^s f(\omega, \omega_c) \, ,
\end{equation}
where $s>0$ is known as Ohmicity parameter, and $\eta>0$ is the coupling strength between the system and the environment. The constant $\omega_c$ is the cut-off frequency, while $f(\omega , \omega_c)$ is the cut-off function, which ensures that $J(\omega) \to 0$ in the limit of large frequencies, i.e. $\omega \to \infty$.  In what follows we consider the exponential cutoff, namely $f (\omega, \omega_c) = e^{-\omega/\omega_c}$.
Depending on the value of $s$, we model different system-environment couplings, corresponding to various physical scenarios~\cite{Leggett_rev:1987,deVega_Alonso_rev:2017,Weiss:2012}. SDs with $s=1$ (i.e. linear in the frequency $\omega$) are called Ohmic, while those for which $s > 1$ ($s < 1$) are known as super-Ohmic (sub-Ohmic).

\begin{figure}
\centering\includegraphics[width=0.9 \columnwidth]{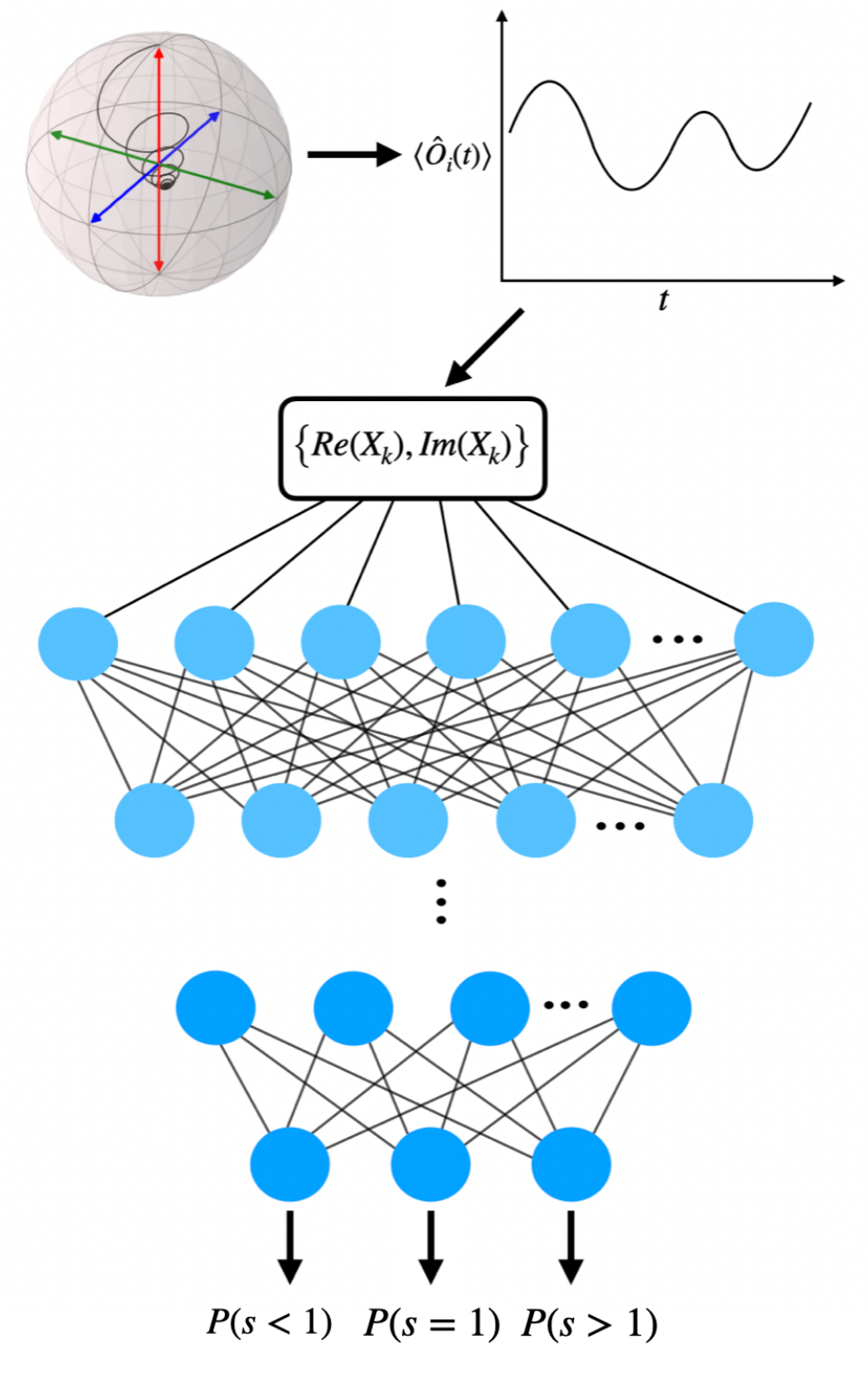}
	\caption{Schematic of the setup: given the time evolution of an observable, denoted as $\langle \hat{O}_i (t) \rangle$, we compute the corresponding Fourier coefficients $\left\{ X_k \right\} $. Then, with the aim of determining which class of spectral density is most compatible with the observed dynamics, we input the real and imaginary parts of the coefficients to a Neural Network. The outputs of the three artifical neurons in the output layer are the probabilities that the input belongs to each of the classes.} \label{fig:procedure}
\end{figure}

In this work, we will use the tools provided by ML  to classify the SD characterising the system-environment interaction.
Specifically, we use an artificial NN that comprises many artificial neurons -- essentially a computational unit -- arranged in a series of layers, as in Fig.~\ref{fig:procedure}~\cite{Marquardt:2021,Marquardt:2023}. Given a set of inputs $\{x_i\}$,  each neuron computes the weighted sum 
\begin{equation}
    z = \sum_i w_i x_i + b \, ,
\end{equation}
with weights $w_i$ and a bias term $b$. A non-linear activation function $f$ is then applied to the result $z$, yielding the output of the neuron $y=f(z)$. The activation function used in this work is the standard sigmoid function, i.e. $f(z) = 1/(1+e^{-z})$. The aforementioned weights and biases are free parameters to be optimised. In addition, the outputs from each layer are input to the next layer. In this way, the input data propagates through the network, so that outputs from later layers become increasingly complex functions of the data. The first layer receives the input data and passes it to the subsequent layer, without performing any computation, while the final layer computes the final output of the network. Accordingly, we refer to these layers as the \emph{input layer} and \emph{the output layer}, respectively. The layers between the input and output layers are known as \emph{hidden layers}. Note that we opt for the aforementioned architecture due to its success in accomplishing the intended objective, without necessitating the use of a more complex architecture, such as a recurrent neural network \cite{Marquardt:2021}. 

For the purpose of classifying the SD using ML techniques, let us suppose we have the time evolution of a family of system observables  $\langle \hat{O}_j (t) \rangle$ (for a set of indices $j$) as input. These time signals can be gathered as outcomes of an experiment carried out in a laboratory, or, as in our case, they can be generated by solving the system dynamics (either exactly or approximately). 

As each signal is a time series, we Fourier-decompose the signal. To this end, we compute the Fourier coefficients as 
\begin{equation}
\label{eq:Fourier1}
     X_k^j  = \sum_{n=0}^{N-1} \langle \hat{O}_j (t_n) \rangle e^{- 2  \pi i k n /N } \, ,
\end{equation}
where $N$ is the total number of time-steps and $\langle \hat{O}_j (t_n) \rangle$ denotes the $n$-th sampled point. We can reconstruct the original signal by inverting \Cref{eq:Fourier1}, where, $X_k^j \in\mathbb{C}$, and the sum runs over all the sampled points in the time series, and $k\in[0,N-1]$.
We split each coefficient $X_k^j$ into its real and imaginary parts and train the network using the Fourier coefficients rather than the time series $\langle \hat{O}_j (t) \rangle$ directly. Using the resulting dataset, we address the ternary classification problem of distinguishing between three different families (i.e. \emph{classes}) of SDs according to their value of the Ohmicity parameter 
[cf. \Cref{eq:Jexp}]. In our case, the output layer of the NN has three artificial neurons which compute weighted sums $z_j$ and apply the softmax activation function \cite{hastie2009elements}, defined as
\begin{equation}
    f(z_j) = \frac{e^{z_j}}{\sum_{j=1}^{N_c} e^{z_j}} \, ,
\end{equation}
where $N_c$ is the number of classes (in our case $N_c=3$). It follows that the outputs of the network are the predicted probabilities that the input belongs to a particular class. As is common for classification problems, we use the categorical cross-entropy as a loss function. Given a dataset containing $N_t$ trajectories, let $y_{ij}$ represent the true probability that the $i$-th trajectory belongs to the $j$-th class and let $\hat{y}_{ij}$ denote the predicted probability of the same. Then the categorical cross-entropy is defined as \cite{murphy2012machine3}

\begin{equation}
    L(\hat{y}, y) = - \frac{1}{N_t} \sum_{i=1}^{N_t} \sum_{j=1}^{N_c} y_{ij} \log (\hat{y}_{ij}) \, .
\end{equation}
The task of training the network reduces to an optimisation problem where the aim is to find the set of parameters that minimises the loss function. A schematic view of the setup is shown in \Cref{fig:procedure}.

\section{Generation of the dataset: spin-boson models}
\label{sec:models}

Given the general framework outlined in \Cref{sec:methods},  we now identify  the systems to scrutinise.
We focus on the dynamics of a Spin-Boson (SB) model consisting of a two-level system interacting with a bosonic bath. Therefore, in \Cref{eq:Hamiltonian}, we choose
\begin{align}
\label{eq:HS_TLS}
\hat{H}_S = \frac{\omega_0}{2} \hat{\sigma}_z \, ,\, \hat{H}_B = \sum_k \omega_k \hat{b}_k^\dagger \hat{b}_k
\end{align}
with $\hat{\sigma}_z$ being the $z$ Pauli operator. The choice of the system-environment coupling Hamiltonian $\hat{H}_I$
leads to different physical scenarios, in general requiring different techniques to solve the dynamics. In \Cref{subsec:exactlysolvable}, we introduce an exactly solvable SB model, where the full system-environment unitary dynamics can be accessed, and the system dynamics is obtained by  tracing out the environmental degrees of freedom. In \Cref{subsec:nonMarkovSB} we then choose a different form of coupling, which requires further approximations to effectively trace out the environment.

In both cases, the reduced dynamics of the system is governed by a master equation of the form
\begin{align}
\label{eq:Liouville}
    \dot{\hat{\rho}} = \mathcal{L}_t \hat{\rho} \, ,
\end{align}
where $\mathcal{L}_t$ is the Liouvillian (super)-operator accounting for both the unitary and non-unitary dynamics, and $\hat{\rho}$ is the reduced density operator. Given the initial state of the system $\hat{\rho}(0) = \hat{\rho}^{0}$,  \Cref{eq:Liouville} can be formally solved yielding $\hat{\rho} = \hat{\rho}(t) = e^{\mathcal{L}_t t} \hat{\rho}^0$ at 
any time $t$. It is thus immediate to obtain the expectation value of a generic observable $\hat{O}$, i.e. $\langle \hat{O} (t) \rangle \equiv \operatorname{tr}_S \left ( \hat{O} \hat{\rho}(t) \right)$. Since we are considering a SB model, a natural choice of the observable would be given by the Pauli operators, i.e. $(\hat{O}_1,\hat{O}_2,\hat{O}_3) = (\hat{\sigma}_x,\hat{\sigma}_y,\hat{\sigma}_z)$ or a combination thereof.

\subsection{Pure Dephasing}
\label{subsec:exactlysolvable}

Let us consider the case in which $\hat{X} = \hat{\sigma}_z$ in \Cref{eq:H_I}. Owing to this choice, the interaction Hamiltonian commutes with the system Hamiltonian and 
the populations of the reduced density matrix are left invariant by the dynamics. In this case, we can access the full unitary evolution, and exactly trace out the environmental degrees of freedom, thus yielding an analytical solution for the reduced dynamics~\cite{Breuer-Petruccione1, Palma:96, Guarnieri:2014}. In \Cref{app:B}, we explicitly solve the dynamics under the standard assumption of an initially uncorrelated system-environment state, where we assume the environment to be in a thermal Gibbs state. Working in the interaction picture, the evolved reduced density matrix at time $t$ can be written in the $\hat{\sigma}_z$ basis $\{ \ket{0}, \ket{1} \}$ as
\begin{equation} \label{eq:Exactlysolvablerhot}
    \hat{\rho}(t) = \begin{pmatrix}
    \rho_{00}^0 & \rho_{01}^0  e^{- \Gamma (t)} \\
    \rho_{01}^{0*} e^{- \Gamma (t)} & 1-\rho_{00}^0
    \end{pmatrix},
\end{equation}
with the decoherence function 
\begin{equation}\label{eq:decoherencefunction}
    \Gamma (t) = 4 \int_0^{\infty} \textrm{d} \omega J (\omega) \coth \left( \frac{\beta \omega}{2} \right) \frac{1 - \cos (\omega t)}{\omega^2} \, .
\end{equation}
From \Cref{eq:Exactlysolvablerhot} we can easily deduce that the interaction with the environment induces pure dephasing in the $\hat{\sigma}_z$ basis, with no dissipation (as deduced by comparing
 \Cref{eq:decoherencefunction} with 
 \Cref{eq:bosonic_corr_f2}). Moreover, it is worth emphasising that there might be choices of the SD leading to negative values of $\Gamma(t)$. In such intervals of time, the system \emph{re}-coheres as a result of 
(non-Markovian) memory effects of the dynamics~\cite{Addis:2014}. 

\subsection{Amplitude Damping}
\label{subsec:nonMarkovSB}

 Alternatively, we can turn to a set-up beyond pure dephasing, just by choosing $\hat{X} = - \hat{\sigma}_x /2$ in the interaction Hamiltonian of \Cref{eq:H_I}. Unlike the case discussed in \Cref{subsec:exactlysolvable}, the Hamiltonian does not exhibit any explicit symmetry, therefore we are not able to provide an exact solution for the dynamics. We can nevertheless effectively solve the dynamics, provided that we rely on further assumptions. Starting from an initial uncorrelated state, we can derive a master equation in the weak coupling regime, where we are still able to obtain non-Markovian effects. As outlined in the \Cref{app:C}, we can derive a second-order approximated master equation that is local in time~\cite{Breuer-Petruccione1,Breuer:2001,Clos:2012} and can be written in terms of dynamical equations for the components of the Bloch vector $\langle \vec{\sigma} (t)\rangle  = \left( \langle \hat{\sigma}_x (t) \rangle, \langle \hat{\sigma}_y (t) \rangle, \langle \hat{\sigma}_z (t) \rangle  \right)^{\rm T}$, with $\langle \hat{\sigma}_i \rangle = {\rm tr}_S (\hat{\sigma}_i \hat{\rho})$. These equations can be cast in the form
\begin{equation}\label{eq:blochME}
    \frac{ \textrm{d} \langle \vec{\sigma} (t) \rangle }{\textrm{d} t} = A(t)  \langle \vec{\sigma} (t) \rangle  + \vec{b} (t)
\end{equation}
with $\vec{b} (t) = \left(0, \, 0, \, b_z (t) \right)^T$ and $b_z (t) = \int_0^t \textrm{d} s \, \mu (s) \sin \left( \omega_0 s \right)$. We have also introduced the matrix 
 \begin{equation}\label{eq:A}
     A (t) = \begin{pmatrix} 
     0 & - \omega_0 & 0 \\
     \omega_0 + a_{yx} (t) & a_{zz} (t) & 0 \\
     0 & 0 & a_{zz} (t)
     \end{pmatrix} 
 \end{equation}
with the time-dependent entries
 \begin{align}
 \label{eq:time_dep_coeff1}
     a_{yx} (t) & = \int_0^t \textrm{d} s \, \nu (s) \sin \left( \omega_0 s \right), \\
\label{eq:time_dep_coeff2}
     a_{zz} (t) &= - \int_0^t \textrm{d} s \, \nu (s) \cos \left( \omega_0 s \right).
 \end{align}
The noise and dissipation kernels $\nu(t)$ and $\mu(t)$ are defined in \Cref{eq:bosonic_corr_f2}.

\section{Analysis and Results}
\label{sec:results}

In this Section, we present the results of our numerical experiments. We consider a two-level system, whose open dynamics depends on the choice of the coupling between the system and the bosonic environment, as discussed in \Cref{sec:models}. For a given initial state, we generate a set of curves reproducing the time evolution of the expectation value of a system observable, i.e. $\langle \hat{O}(t) \rangle$. Each signal is sampled at N = 400 successive and equally spaced points over a certain time interval $[t_{\rm min}, t_{\rm max}]$, to ensure a sufficient resolution of the dynamics. As discussed in \Cref{sec:methods}, instead of directly using the time series, we input the $2 N$ real and imaginary parts of the Fourier coefficients $X_k$. For this reason, we build the input layer with $2 N$ input neurons. The NN for each model consists of the input layer followed by $2$ hidden layers where the first hidden layer comprises 250 neurons, and the second comprises $80$ neurons. The output layer, instead, is made of $3$ neurons, which matches the number of classes (Ohmic, sub-Ohmic, super-Ohmic). The choice of network architecture was iteratively refined, adding layers and neurons until the network achieved a high accuracy without overfitting. The code employed for data generation, the datasets, and the code utilised for subsequent analysis are available in the following GitHub respository \cite{github3}.

In order to evaluate the performance of the NN, we use the classification accuracy which is defined as the percentage of trajectories that are classified correctly. We generate a training dataset containing $N_{\text{Train}}$ trajectories which is used to train the model, a validation dataset containing $N_{\text{Valid}}$ trajectories which is used to assess the performance during training, and a test dataset containing $N_{\textit{Test}}$ trajectories which is used to assess the final accuracy of the network. We optimise the NNs using whole batch gradient descent and the Adam optimiser with a learning rate of $1 \times 10^{-4}$.

\subsection{Pure Dephasing}

\begin{figure*}
\centering
\subfloat[]{\label{fig:trainingset_separated}{\includegraphics[width=0.5\textwidth]{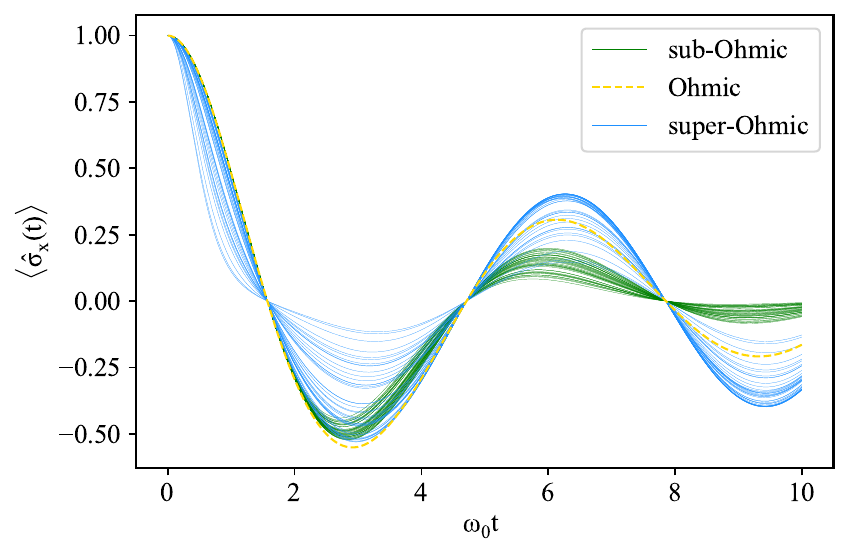}}}
\subfloat[]{\label{fig:trainingset_notseparated}{\includegraphics[width=0.5\textwidth]{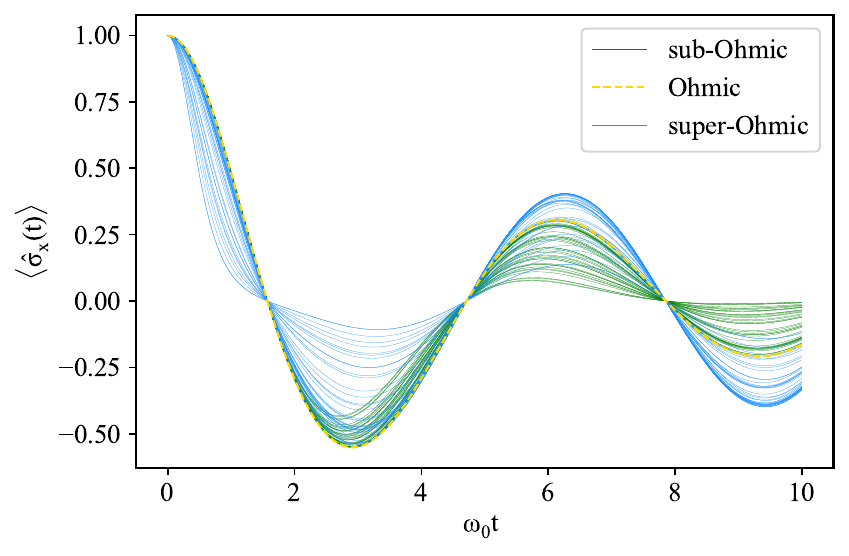}}}
\hfill
\subfloat[]{\label{fig:trainingset_0.25and0.25}{\includegraphics[width=0.5\textwidth]{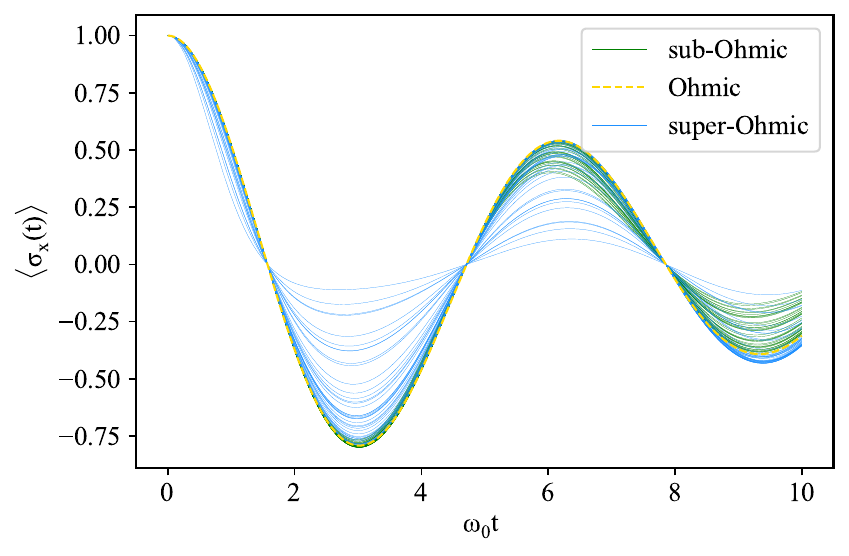}}}
\subfloat[]{\label{fig:trainingset_0.25and2.05}{\includegraphics[width=0.5\textwidth]{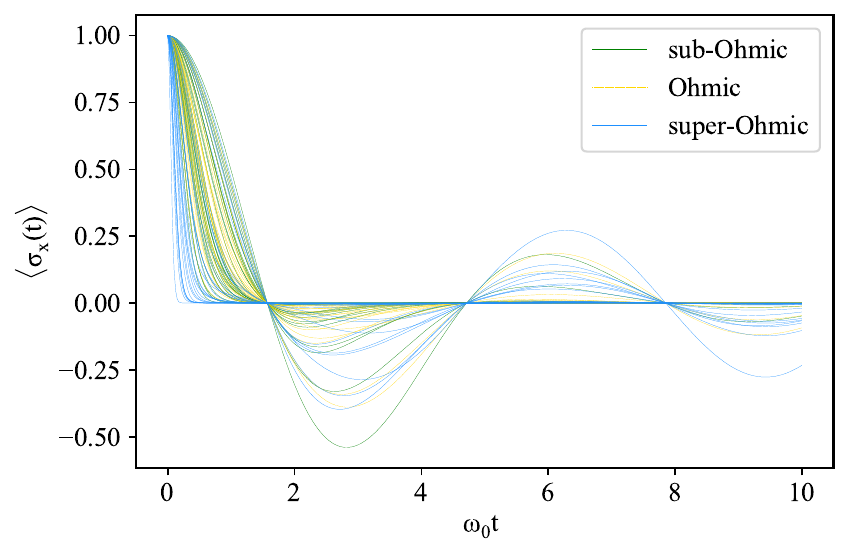}}}
\caption{Pure dephasing model: some of the curves from the datasets used to train the NN. Each curve represents the time evolution of the observable  $\langle \hat{\sigma}_x(t)\rangle$ for the initial state $\hat{\rho}^0 = \ket{+}\!\bra{+}$, with $\beta \to \infty$. The curves shown in panels (a) and (b) are generated by choosing $\eta = 0.25$ and $\omega_c = 0.5$. In panel (a) we have taken $s \in (0, 0.5]$ ($s \in [1.5, 4]$) if the SD is sub-Ohmic (super-Ohmic). The curves in panels (b), (c) and (d) are generated by choosing $s \in (0,1)$ ($s \in (1, 4]$) if the SD is sub-Ohmic (super-Ohmic). In panel (c) we have taken $\eta = \omega_c = 0.25$, whereas $\eta, \omega_c \in [0.25, 2.05]$ in panel (d). The green curves in each panel correspond to sub-Ohmic dissipation while the yellow and blue correspond to Ohmic and super-Ohmic dissipation, respectively.}
\label{fig:subfigures}
\end{figure*}

We consider the evolution of the pure dephasing model introduced in \Cref{subsec:exactlysolvable}.  We solve the system dynamics choosing the initial state $\hat{\rho}^0 = \ket{+}\!\bra{+}$, with $\ket{+} = (\ket{0} + \ket{1})/{\sqrt{2}}$, while --- without loss of generality --- we keep the thermal bath at zero temperature, i.e. $\beta \to \infty$. With this choice, we obtain the expectation value $\langle \hat{\sigma}_x (t) \rangle$ within the time interval $t\in[0,10]$.  It is worth noting that alternative choices for the initial state and the observable can be made, however, it should be recognised that, within the context of the pure dephasing model, the time evolution of $\langle \hat{\sigma}_z (t) \rangle$ will always be trivial. Moreover, should the initial state possess coherences equal to zero, the time evolution of the density matrix, and by extension any observables, will be trivial as well. As input to the NN we use the real and imaginary components of the Fourier coefficients obtained using \Cref{eq:Fourier1}. We generate a training, validation, and test set of size $N_{\text{Train}}/2=N_{\text{Valid}}=N_{\text{Test}}=2400$. The number of trajectories in the Ohmic, sub-Ohmic and super-Ohmic classes are equal in all datasets.

We assess the performance of the NN in two scenarios: the first being where $\omega_c$ and $\eta$ are fixed, and the second being where they vary. In the first scenario, we consider the case where $\eta = 0.25$, $\omega_c = 0.5$, while the only parameter that varies is $s$. At first, we want to test how the model performs when the classes are easy to differentiate. To that end, we consider trajectories with  $s \in \left( 0, 0.5 \right]$ if the SD is sub-Ohmic and $s \in \left[ 1.5, 4 \right]$ if it is super-Ohmic. If the SD is Ohmic then s = 1. In Figure \ref{fig:trainingset_separated} a subset of trajectories from the resulting training set are plotted where the green curves correspond to sub-Ohmic dissipation, while the yellow and blue curves are trajectories characterised by Ohmic and super-Ohmic dissipation, respectively. Given the substantial separation in the permissible values for $s$ across the different classes, we expect that the performance of the NN will be high. In Figure \ref{fig:trainingset_separated}, it is evident that the classes are easily distinguishable due to distinct characteristics exhibited by each of them. Specifically, the super-Ohmic curves exhibit the steepest initial descent. In addition, while the sub-Ohmic and Ohmic curves show a comparable initial rate of descent, their oscillatory patterns differ. Oscillations are exhibited by all three classes, but the amplitude of oscillation for the sub-Ohmic curves appears to reduce more rapidly than that of the Ohmic or super-Ohmic curves as time grows. Confirming our expectation, the accuracy of the network evaluated on both the training and the test set reaches $100\%$ after  $\approx 80$ training iterations. 

We then make the task a bit more difficult for the network by allowing $s \in (0, 1)$ for the sub-Ohmic dissipation and $s \in (1,4]$ for the super-Ohmic dissipation. We anticipate that the task will be more difficult in this scenario due to the reduced separation in the allowed values for $s$ across the classes. This is reflected in the resulting training trajectories, a subset of which are plotted in \Cref{fig:trainingset_notseparated}, where we observe that the super-Ohmic curves maintain a more pronounced initial descent relative to the Ohmic and sub-Ohmic curves. However, there are instances where the oscillation amplitudes between the classes are similar. The final training accuracy of the network in this case reaches $99.31 \% $ after around $5000$ training iterations, while the final test accuracy reaches $99.50 \%$.

\begin{figure}
\centering\includegraphics[width=\columnwidth]{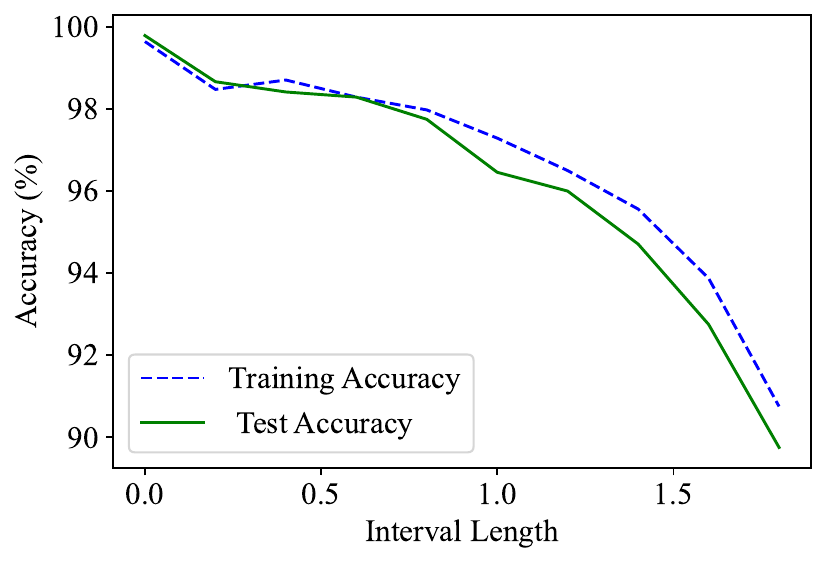}
	\caption{Pure dephasing model: the classification accuracy against the length of the interval from which $\eta$ and $\omega_c$ are sampled.} \label{fig:accvintervallength}
\end{figure}

Next, to challenge the NN further, we consider the second scenario where we let $\eta$ and $\omega_c$ vary: the idea is to assess the performance as we increase the upper bounds of the intervals from which they are sampled. We let $s \in \left(0, 1 \right)$ for the sub-Ohmic spectral densities and $s \in \left(1, 4 \right]$ for the super-Ohmic spectral densities. Initially, we set both $\eta$ and $\omega_c$ equal to $0.25$, then we let them vary into the interval $\left[ 0.25, 0.45 \right]$. We increase the upper bound in increments of $0.2$ until the interval becomes $\left[ 0.25, 2.05 \right]$. Figure \ref{fig:trainingset_0.25and0.25} shows some example trajectories from the training set for $\eta = \omega_c = 0.25$, while Figure \ref{fig:trainingset_0.25and2.05} shows some for $\eta, \omega_c \in \left[ 0.25, 2.05 \right]$. From Figure \ref{fig:trainingset_0.25and0.25}, we can observe that the scenario closely resembles that depicted in Figure \ref{fig:trainingset_notseparated}. In particular, the initial decay rates of the super-Ohmic curves are larger than those corresponding to the Ohmic or sub-Ohmic curves. However, the oscillation amplitudes across the three classes are comparable in some cases. In Figure \ref{fig:trainingset_0.25and2.05}, it is evident that there is considerable overlap both in the initial decay rates and the amplitudes of oscillation between the three classes, indicating that the differences between the behaviours of the classes are less pronounced and that the classification task will be significantly more difficult. The classification results after $2 \times 10^{4}$ training iterations are shown in figure \ref{fig:accvintervallength} where the blue curve is the accuracy evaluated on the training set and the green curve is the accuracy evaluated on the test set. As expected, we can see that the accuracy decreases as we consider larger intervals $\eta$ and $\omega_c$. This is indeed the case, as taking larger intervals essentially increases the amount of noise in the dataset. It is worth noting that the accuracy may improve with larger datasets or more training iterations.

\subsubsection{Measurement Sampling Noise}

The accurate measurement and classification of experimental expectation values are inherently impacted by various sources of noise. One of the predominant ones is sampling noise, which arises due to the finite number of measurement samples that one can realistically acquire experimentally. In this Section, we analyse the impact that sampling noise has on the NN, with the aim of providing a deeper insight into the performance of the model under realistic conditions. In our approach, we artificially introduce noise into the trajectories by adding a random value to each time point. Such value is drawn from a normal distribution with zero mean and a given standard deviation, $\sigma$.

\begin{figure*}
\centering
\subfloat[]{\label{fig:accvstd_separated}{\includegraphics[width=0.5\textwidth]{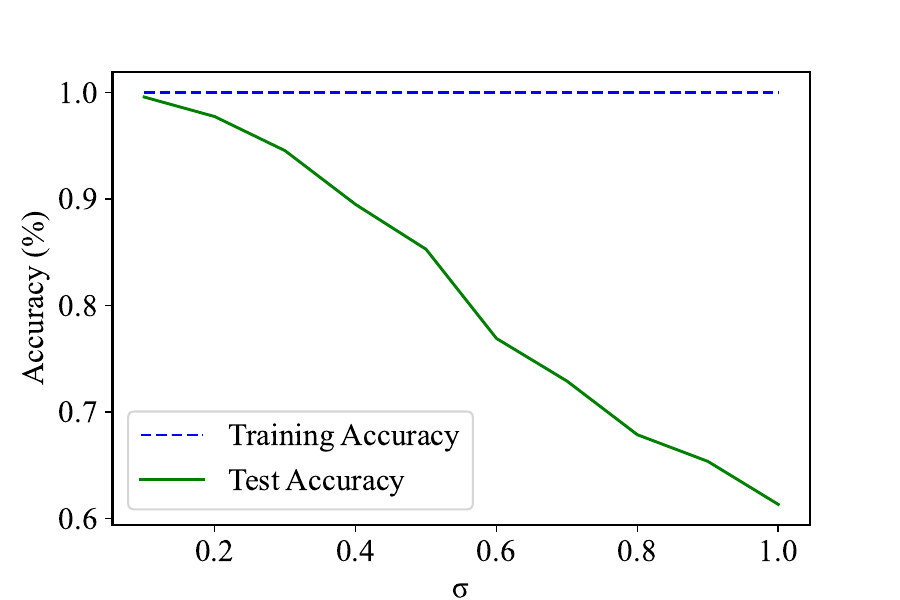}}}
\subfloat[]{\label{fig:accvstd_notseparated}{\includegraphics[width=0.5\textwidth]{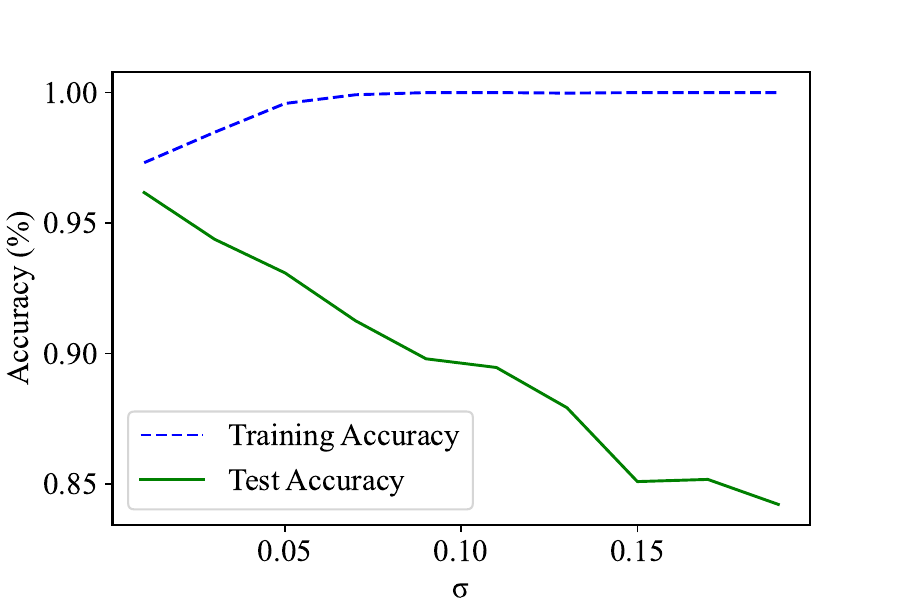}}}
\hfill
\subfloat[]{\label{fig:accvstd_0.25and0.25}{\includegraphics[width=0.5\textwidth]{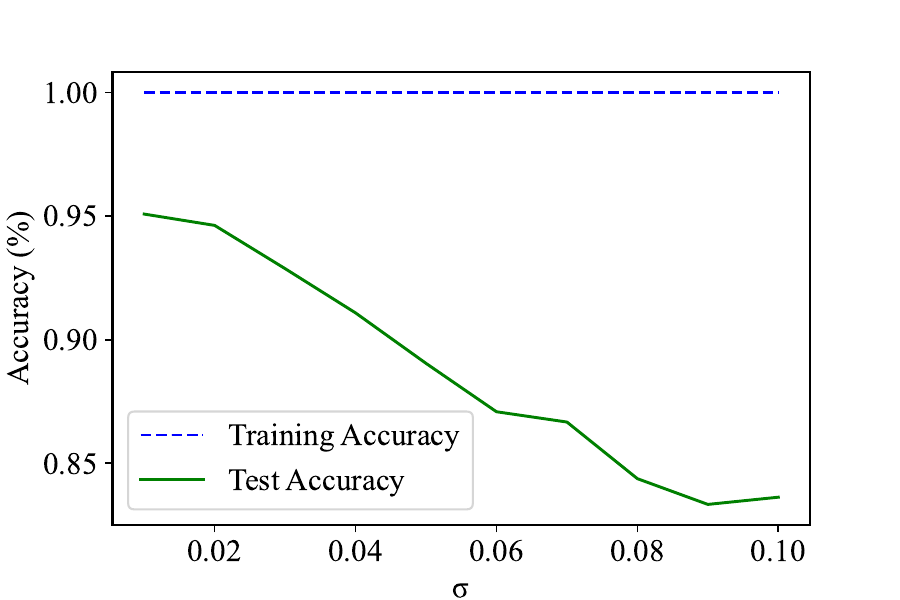}}}
\subfloat[]{\label{fig:accvstd_0.25and2.05}{\includegraphics[width=0.5\textwidth]{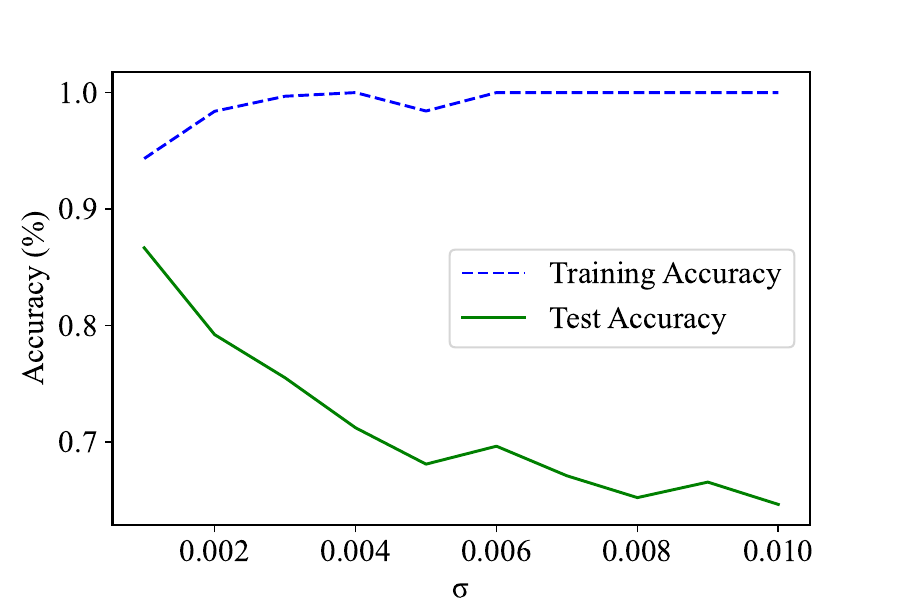}}}
\caption{ Pure dephasing model: The training and test accuracy of the NN in relation to the standard deviation $\sigma$ of the artificial noise. Panels (a) and (b) show results for datasets with $\eta = 0.25$ and $\omega_c = 0.5$. In panel (a) we have $s \in (0, 0.5]$ for sub-Ohmic SDs and $s \in [1.5,4]$ for super-Ohmic SDs. The datasets in panels (b), (c) and (d) are characterised by $s \in (0,1)$ for sub-Ohmic SDs and $s \in (1,4]$ for super-Ohmic SDs. In panel (c) the parameters are set as $\eta = \omega_c = 0.25$, whereas in panel (d) we have $\eta , \omega_c \in [0.25, 2.05]$. The blue dashed lines represent the training accuracy, while the green solid lines show the test accuracy. }
\label{fig:accvstd}
\end{figure*}

We assess how the performance of the NN varies with $\sigma$ in the same two scenarios as before: firstly, we hold both $\eta = 0.25$ and $\omega_c = 0.5$ constant. For the dataset with clear separation in the allowed values of $s$ among classes [$s \in (0, 0.5)$ for a sub-Ohmic SD; $s \in [1.5, 4]$ for a super-Ohmic one], the results after $10^3$ training iterations are shown in Figure~\ref{fig:accvstd_separated}. The training accuracy remains consistently close to $100 \%$ for all of the considered $\sigma$ values. This suggests that the NN can learn from the training data well, regardless of the magnitude of the noise that is introduced. However, the test accuracy decreases from $99.58 \%$ for $\sigma = 0.1$ to $61.33 \%$ for $\sigma = 1$, thus indicating that the capacity of the model to generalise to unseen data diminishes as the noise intensity increases.

For the dataset with $\eta = 0.25$, $\omega_c = 0.5$  and $s \in (0 , 1)$ -- for a sub-Ohmic SD -- and $s \in (1,4]$ -- for a super-Ohmic one -- the results after $10^3$ training iterations are shown in Figure~\ref{fig:accvstd_notseparated}. Mirroring the trends observed for the preceding dataset, the training accuracy remains notably high and close to $100 \%$ for the range of $\sigma$ examined. On the other hand, the test accuracy starts at $96.17 \%$ for $\sigma = 0.01$ and decreases to $84.21 \%$ for $\sigma = 0.19$. Therefore, 
relative to the previously examined case, the NNs performance with this dataset exhibits a heightened susceptibility to noise. 

We now redirect our attention to the case where $\eta$ and $\omega_c$ vary. To begin with, we analyse the dataset corresponding to the shortest interval length in Figure \ref{fig:accvintervallength} [where $s \in (0,1)$ if the SD is sub-Ohmic and $s \in (1,4]$ if it is super-Ohmic] with $\eta = \omega_c = 0.25$. The results of this analysis after $10^4$ training iterations, shown in Figure \ref{fig:accvstd_0.25and0.25}, closely resemble those in Figure \ref{fig:accvstd_notseparated}, albeit with a noticeable decrease in performance. The training accuracy remains at $100 \%$ while the test accuracy starts at $95.08 \%$ for $\sigma = 0.01$ and decreases to $83.63 \%$ for $\sigma = 0.1$. Lastly, we turn our attention to the dataset corresponding to the longest interval length in Figure \ref{fig:accvintervallength}. The results after $2 \times 10^4$ training iterations are shown in Figure \ref{fig:accvstd_0.25and2.05}. While the conditions for $s$ are consistent with those defined for the shortest interval length, $\eta$ and $\omega_c$ vary into the interval $[0.25, 2.05]$. The training accuracy for this dataset starts at $94.33\%$ for $\sigma = 0.001$ and exhibits minor fluctuations across the considered $\sigma$ range but remains quite close to $100 \%$. Meanwhile, the test accuracy begins at $86.67 \%$ for $\sigma = 0.001$ and drops to $64.63 \%$ at $\sigma = 0.01$. This dataset exhibits the greatest sensitivity to noise, leading to the lowest performance metrics. Moreover, the results emphasise the fact that despite the NNs ability to learn from training data, increasing noise levels hamper its generalisation to previously unseen data.

\subsection{Amplitude Damping}

\begin{figure}
\centering\includegraphics[width=\columnwidth]{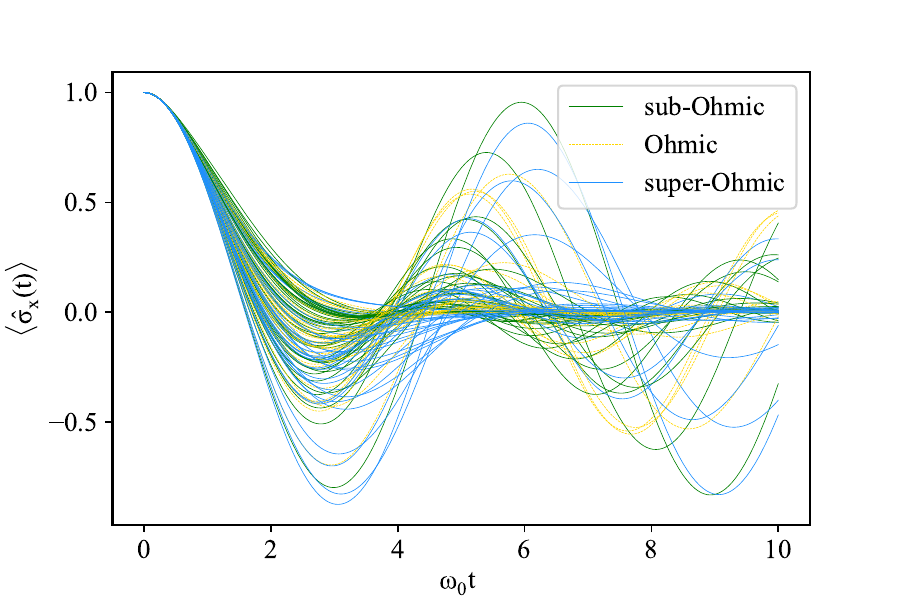}
	\caption{Amplitude damping model: some example curves from the training set. Each curve represents the time evolution of the observable $\langle \hat{\sigma}_x (t) \rangle$ for the initial state $\hat{\rho}^0 = \ket{+}\bra{+}$, where $\beta = 0.1$, $\eta \in (0, 0.2]$, $\omega_c \in [0.1, 2]$. We choose $s \in (1, 2]$ if the spectral density is super-Ohmic, and $s \in [0.3, 1)$ if the spectral density is sub-Ohmic. The green curves correspond to sub-Ohmic dissipation while the yellow and blue correspond to Ohmic and super-Ohmic dissipation, respectively.} \label{fig:trainingsetNM}
\end{figure}

We shall now analyse the amplitude damping model detailed in section~\ref{subsec:nonMarkovSB}. We choose the initial state $\hat{\rho}^0 = \ket{+}\!\bra{+}$, the bare frequency of the oscillator $\omega_0 = 1$, while we keep the environmental inverse temperature $\beta = 0.1$. We subsequently solve for the dynamics of the system and determine the expectation value $\langle \hat{\sigma}_x (t) \rangle$ within the time interval $t \in [0, 10]$. As for the previous model, we use the real and imaginary components of the Fourier coefficients obtained through \Cref{eq:Fourier1} as input to the NN. We let $\eta \in (0, 0.2]$, $\omega_c \in [0.1, 2]$. In addition, we take $s \in (1, 2]$ [$s \in [0.3, 1)$] if the SD is super-Ohmic [sub-Ohmic] and $s=1$ if the SD is Ohmic.
We generate a training, validation, and test set such that $N_{\text{Train}} = 1500$, and $N_{\text{Valid}} = N_{\text{Test}} = 300$. In all datasets, the Ohmic, sub-Ohmic and super-Ohmic classes have an equal number of trajectories. Figure \ref{fig:trainingsetNM} shows some of the curves from the resulting training set where, as before, the green curves represent sub-Ohmic dissipation while the yellow and blue curves correspond to trajectories characterised by Ohmic and super-Ohmic dissipation, respectively. The final training accuracy of the network in this case reaches $97.93 \%$ after $10^4$ training iterations while the test accuracy is significantly lower and reaches $93.00 \%$.

Firstly, we would like to assess the number of  time-points required to attain a high level of accuracy. It should be noted that it is generally advisable to avoid having highly correlated features in a dataset, whose 
linear dependence implies that the value of one can be derived from that of the other \cite{hastie2009elements}. Hence, mutually correlated features convey redundant information to the model since each feature provides little or no additional information beyond what the other features already capture. Including all of the features will not improve the ability of the model to discriminate, but will increase the complexity of the algorithm, thus increasing the
computational cost. 

\begin{figure}
\centering\includegraphics[width=\columnwidth]{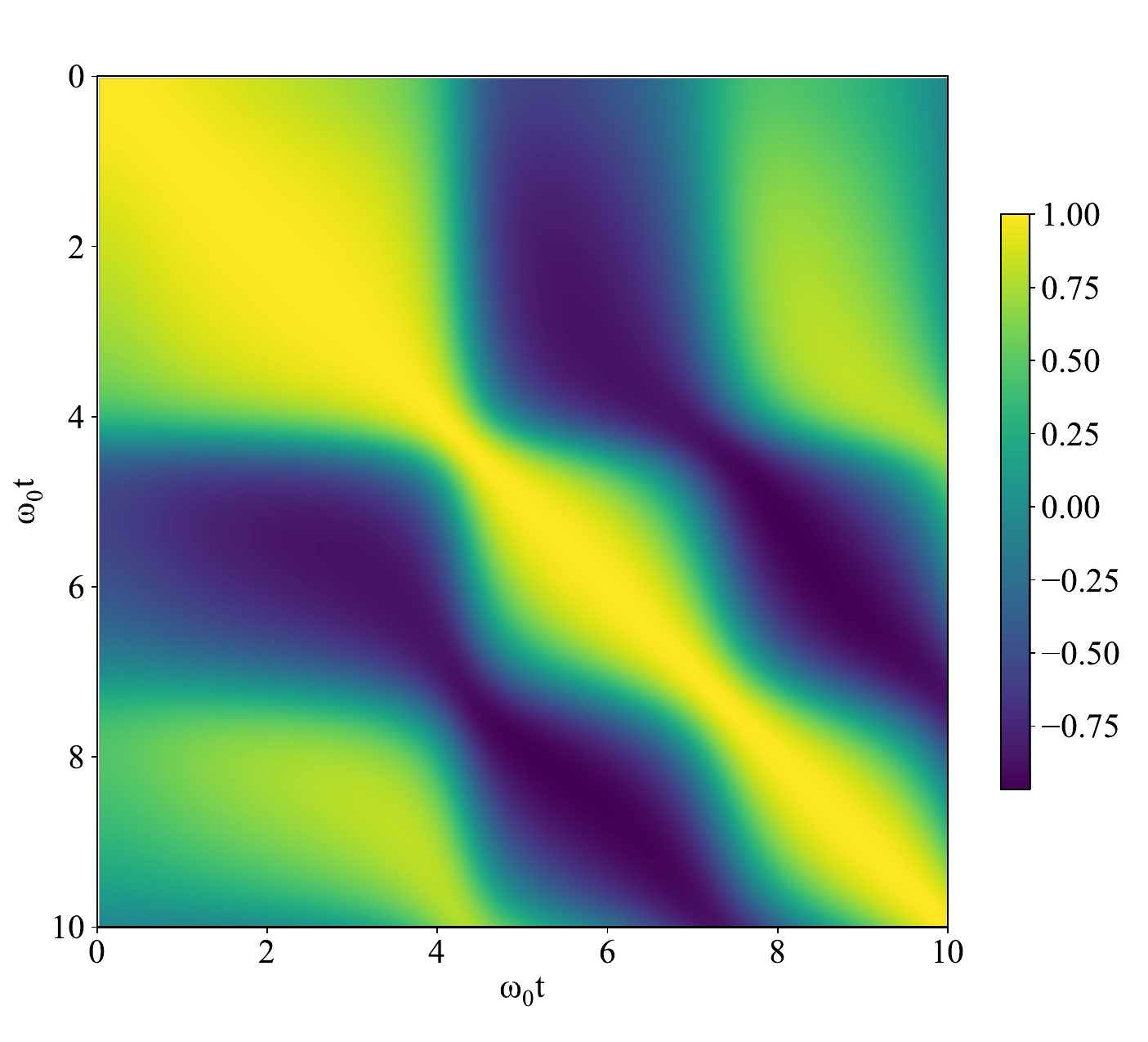}
	\caption{Amplitude damping model: The correlation heatmap for the entries $C_{ij}$ of the correlation matrix $\mathbf{C}$, where $C_{ij}$ is the Pearson Correlation coefficient between the $i$-th and $j$-th time-point in the training set [cf. \Cref{eq:Pearson}].} \label{fig:corrmatrix}
\end{figure}

To this end, we introduce the Pearson correlation coefficient, which is a statistical measure of linear correlation between two variables~\cite{wilcox2010fundamentals3, james2013introduction2}. It ranges from a value of $-1$, indicating perfect anti-correlations, to $1$, when the variables are perfectly correlated. A value of $0$ indicates that there is no linear relationship between the two variables. Let  $\langle \hat{\sigma}_x \rangle_n^i$ denote the $n$-th time-point of the $i$-th trajectory in a given dataset. Then the Pearson correlation coefficient between the $n$-th and $m$-th time-points, denoted as $C_{nm}$, is given by the formula
\begin{equation}
\label{eq:Pearson}
    C_{nm} \equiv \frac{ \sum_{i=1}^{N} \Delta \langle{\hat{\sigma}_x} \rangle^i_n \ \Delta \langle{\hat{\sigma}_x} \rangle^i_m}{\sqrt{\sum_{i=1}^{N} \left ( \Delta \langle{\hat{\sigma}_x} \rangle^i_n\right)^2}\sqrt{\sum_{i=1}^{N} \left ( \Delta \langle{\hat{\sigma}_x} \rangle^i_m \right )^2}} \, , 
\end{equation}
where $\Delta \langle{\hat{\sigma}_x} \rangle^i_j \equiv \langle \hat{\sigma}_x  \rangle_j^i - \overline{ \langle  \hat{\sigma}_x \rangle}_j$, with $\overline{ \langle  \sigma_x\rangle}_n $ the average value of the $n$-th time step, and $N$ the total number of trajectories in the dataset. We calculate the Pearson correlation coefficient between each time step in our training set and generate a correlation matrix, $\boldsymbol{C}$, whose entries quantify the correlation between time-points. The resulting correlation heatmap, a graphical representation of the correlation matrix, is shown in~\Cref{fig:corrmatrix}. From the heatmap, it can be observed that there is a high degree of correlation between adjacent and near-adjacent time points.

To address this issue, a common approach is to perform \emph{feature selection}, identifying a subset of features that are the most informative and non-redudant. Retaining only one of two correlated features may expedite the learning process, without compromising the accuracy of the model. While we ideally want to avoid correlation between the features in a dataset, it is preferable to retain features which are correlated with the dependent variable  \cite{liu2012feature3}. Correlations make it possible to use the value of one variable to predict the value of another, meaning that features which are correlated with the output are predictive of the output. 

Note that the Pearson correlation coefficient is only suitable for measuring the correlation between two continuous variables. As, in our case, the dependent variable consists of discrete labels, we can instead determine the degree of correlation between a feature and the dependent variable by examining whether the variance of the feature can be explained by the dependent variable. To do this, we group the feature into classes based on the discrete labels, compute the variance of each class, and calculate the difference between the mean of the resulting variances and the overall variance of the feature. If the mean of the class variances is significantly lower than the overall variance, this suggests that the feature and the dependent variable are correlated.

A possible strategy for performing feature selection and identifying the most salient features for learning is thus to sort the entries in the correlation matrix into descending order. Then, starting from the highest correlation, one can remove the contributing feature that exhibits the lowest correlation with the dependent variable. Using the above strategy, we can obtain a ranking of the features based on their importance and determine the order in which to remove features if we are to maintain a high classification accuracy. In this scenario, given that the time intervals between points may not be uniformly distributed, it becomes necessary to compute the Fourier coefficients using the non-uniform discrete Fourier transform \cite{doi:10.1137/S003614450343200X11}
\begin{equation}
\label{eq:nuFourier}
X_k = \sum_{n=0}^{N-1} \langle \hat{\sigma}_x (t_n) \rangle e^{- 2 \pi i k p_n } \, ,      
\end{equation}                   
where  $p_n$ are the non-uniform time points suitably scaled to fall between $0$ and $1$, while $\langle \hat{\sigma}_x (t_n) \rangle$ denotes the $n$-th sampled point in a given trajectory. As for the discrete Fourier transform, $k$ is the frequency which is an integer number between $0$ and $N-1$. Note that if $p_n = n / N$, then this equation reduces to the discrete Fourier transform shown in~\Cref{eq:Fourier1}.

\begin{figure}
\centering\includegraphics[width=\columnwidth]{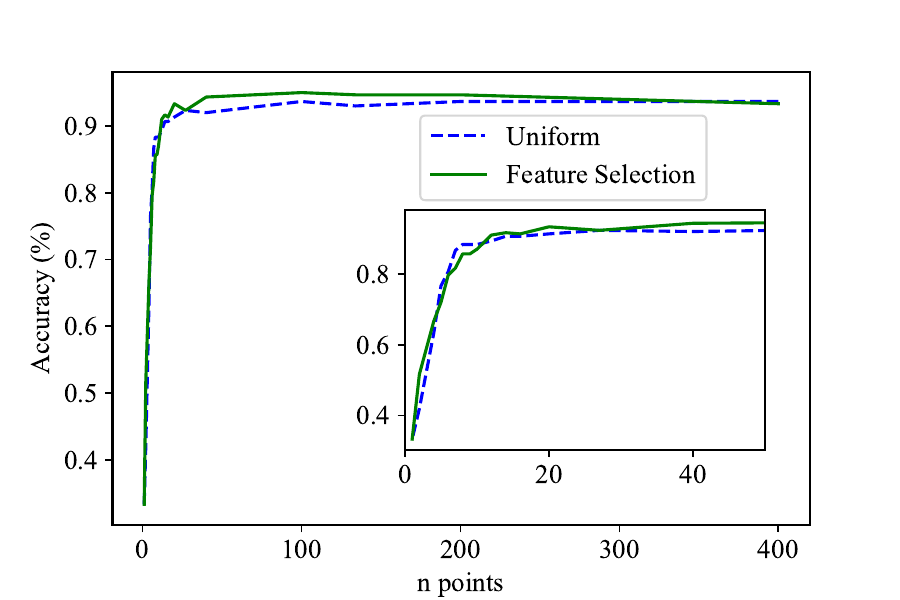}
	\caption{Amplitude damping model: the test accuracy of the NN against the number of time points when the time points are selected uniformly or using the proposed feature selection method described in the main text. } \label{fig:reducepoints}
\end{figure}

We compare the results obtained using the proposed feature selection algorithm with the results obtained by selecting time points uniformly, i.e. choosing time points that are evenly spaced throughout the datasets. For example, we might select the first in every $5$ points or the first in every $100$. \Cref{fig:reducepoints} shows a plot of the test accuracy against the number of selected time points for the two different selection methods. The blue curve show the results obtained using uniform sampling, while the green curve shows the results obtained using the proposed feature selection algorithm. Firstly, the plot shows that the test accuracy remains consistently high until the number of time points is reduced to approximately $20$. Beyond this point, a sharp decline in the accuracy is observed, as shown in the inset of \Cref{fig:reducepoints}. Analysis of the plot indicates that the performance of the two time point selection methods is comparable across different ranges of selected time points. Specifically, when we take a number of time points between $\approx250$ and $\approx400$, there is little difference between the accuracy obtained using uniform sampling and the feature selection algorithm. However, in the range of approximately $40$ to $250$ time points, the feature selection algorithm shows slightly better results compared to uniform sampling. Lastly, taking less than $\approx 40$ points, the test accuracy fluctuates, but we can conclude that the performance of both methods is similar. 

For the sake of completeness, we also explored various other methods for feature selection. For instance, after grouping each feature according to the discrete labels, we used one-way-analysis-of-variance (ANOVA) to determine if there were statistically significant differences between the three groups~\cite{doi:10.4097/kjae.2017.70.1.22222}. We also considered the ratio of the mean of the variances of the groups and the overall variance, as opposed to the difference. Lastly, we attempted to assess the importance of each feature using principal component analysis. Specifically, we examined the degree to which each feature contributed to the  principal components, as a large contribution to the principal components suggests that a feature is important in explaining the overall variability of the data \cite{johnson2002applied1}. We observed that none of the aforementioned methods outperformed the correlation based feature selection algorithm employed in \Cref{fig:reducepoints}.

\subsubsection{Measurement Sampling Noise}

In this section, similar to the analysis conducted for the pure dephasing model, we assess how the NN performs when subject to realistic conditions. Due to the presence of noise sources such as sampling noise, experimentally obtained expectation values are seldom completely accurate, as it is only feasible to collect a finite number of measurement samples experimentally. Given this, we aim to investigate how the performance of the NN is affected by these realistic challenges. To this end, we simulate the impact of sampling noise by incorporating artificial noise into the trajectories. We use the entire trajectory, consisting of $400$ time points, add a random value drawn from a normal distribution with zero mean and a standard deviation $\sigma$ to each point, and then assess the performance of the NN as $\sigma$ increases.

\begin{figure}
\centering\includegraphics[width=\columnwidth]{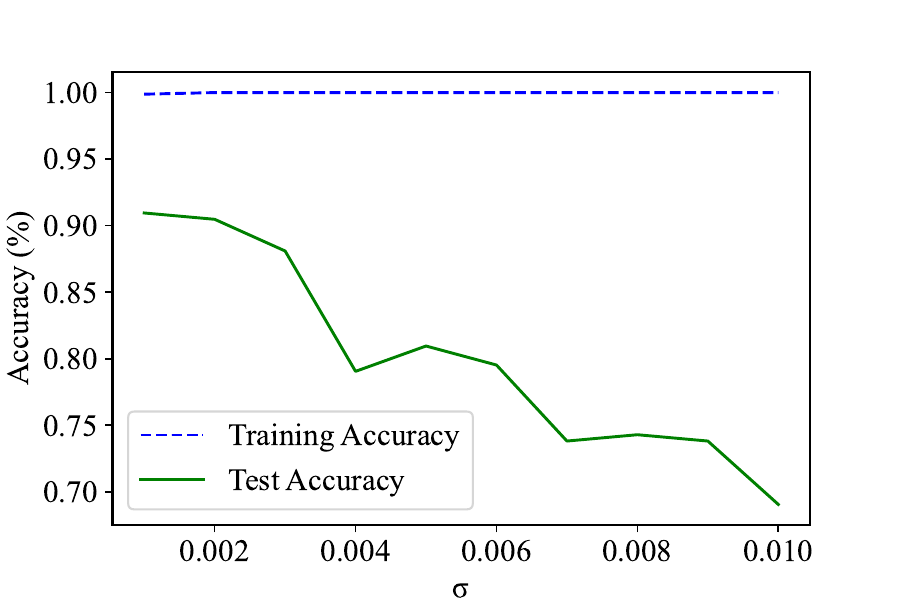}
	\caption{ Amplitude damping model: The training and test accuracy of the NN against the standard deviation, $\sigma$, of the artificial noise.} \label{fig:accvstd_AD}
\end{figure}

The results of our analysis, after $10^4$ training iterations are shown in Figure \ref{fig:accvstd_AD}, where the blue dashed line represents the accuracy evaluated on the training set, and the green solid line corresponds to the accuracy evaluated on the test set. We observe that the training accuracy remains consistently high, at around $100 \%$. In contrast, the test accuracy starts off at $90.95 \%$ when $\sigma = 0.001$ and experiences a decline, dropping to $69.05 \%$ as the value of $\sigma$ increases to $0.01$. Consequently, our observations are consistent with those obtained for the pure dephasing model. The model is able to effectively learn from the training data, and maintain a high training accuracy, regardless of the noise levels. However, its capacity to generalise to new, unseen data deteriorates. 

\section{Conclusions}
\label{sec:conclusions}

We have shown that, in a standard open system scenario, a NN can perform SD-classification with high accuracy. First, we have considered an exactly solvable, pure-dephasing model, and  assessed the performance of the NN as a classifier, highlighting the limiting role played by the fluctuations of the SD parameters. We have then considered a SB model that, under a number of reasonable approximations, results in a master equation accounting for energy losses and decoherence. We observed that, despite the approximations being invoked, the NN can perform the SD-classification task  with high accuracy. Furthermore, we thoroughly discussed the interplay between high accuracy in the classification task and the number of sampled points for the system observable. Lastly, we investigated how the NN's performance for both models withstands the challenge of measurement sampling noise, thus providing insights into its robustness under realistic conditions. 

The methodology introduced in this paper, as well as the case studies analysed therein,  highlight the capability of ML techniques to characterise environments with arbitrary SDs,  thus embodying a reliable tool for environment characterization and the provision of useful information for control and process diagnosis.
This paves the way to, and leaves great hopes for, the full characterization of an unknown SD through, for instance, regression of the parameters rather than classification. 
We also stress that the method put forward here does not rely critically on how the information on the dynamics is specifically acquired. In this sense, we expect the method to maintain effectiveness even when
considering classes of SDs leading to long-lived correlations that, in turn, would hinder the direct derivation of master equations in Lindblad-like form. In such cases, one should rely on
more sophisticated simulation techniques -- such as Hierarchical Equation Of Motion (HEOM)~\cite{Tanimura:1989,Tanimura:1990}, Time-Evolving Matrix Product Operators (TEMPO)~\cite{Lovett:2018}, or Time-Evolving Density with Orthogonal Polynomials Algorithm (TEDOPA)~\cite{Chin:2010,Prior:2010,Tamascelli:2019}, just to name a few. The combinations of one of these methods with the ML will help achieving the successful characterization and control of the environment affecting a given open system.

\acknowledgements
GZ is grateful to Ricardo Puebla for insightful discussions in the early development of the problem. JB and MP thank the Leverhulme Trust Doctoral Scholarship grant LINAS. MP acknowledges the support by the European Union's Horizon 2020 FET-Open project  TEQ (766900),  the Horizon Europe EIC Pathfinder project QuCoM (Grant Agreement No.\,101046973), the Leverhulme Trust Research Project Grant UltraQuTe (grant RGP-2018-266), the Royal Society Wolfson Fellowship (RSWF/R3/183013), the UK EPSRC (EP/T028424/1), and the Department for the Economy Northern Ireland under the US-Ireland R\&D Partnership Programme.

\appendix

\section{The Correlation Function of a Bosonic Bath}
\label{app:A}

Here we explicitly derive the correlation function for an arbitrary quantum system that is interacting with an environment which is made up of infinitely many independent harmonic oscillators, i.e. \Cref{eq:bosonic_corr_f1,eq:bosonic_corr_f2} of the main text. Given an interaction Hamiltonian in the form of \Cref{eq:H_I} and the bath operator $\hat{B}$ given by \Cref{eq:bath_operator}, we can compute the correlation function which is defined as

\begin{equation}
     \alpha_{\beta} (t) = \langle \hat{B}(t) \hat{B}(0) \rangle_B = \operatorname{tr}_B \left(  \hat{B} (t) \hat{B} (0) \hat{\rho}_B \right) \, .
     \label{eq:corr}
\end{equation}

We now move to the interaction picture via the relation $\hat{B}(t) = e^{it \hat{H}_B} \hat{B} e^{-it \hat{H}_B}$, where $\hat{H}_B = \sum_k \omega_k \hat{b}_k^{\dagger} \hat{b}_k$ is the Hamiltonian of a set of independent harmonic oscillators. Therefore, we have

\begin{align}\label{eq:B_int_pic}
    \hat{B} (t) & = \sum_k \left( g_k \hat{b}_k^{\dagger} e^{i \omega_k t} + g_k^* \hat{b}_k e^{-i \omega_k t} \right) \, ,
\\
    \hat{B} (0) &= \sum_k \left( g_k \hat{b}_k^{\dagger} + g_k^* \hat{b}_k \right) \, .
\end{align}

Thus, the expression for the correlation function reads 
\begin{equation}
     \langle \hat{B} (t) \hat{B} (0) \rangle_B = \sum_k | g_k |^2 \left(  \langle \hat{b}_k^{\dagger} \hat{b}_k \rangle_B \, e^{i \omega_k t} + \langle \hat{b}_k \hat{b}_k^{\dagger}  \rangle_B  \,e^{-i \omega_k t}  \right) \, ,
\end{equation}
where we have utilised the fact that $\langle \hat{b}_k \hat{b}_l \rangle_B = \langle \hat{b}_k^{\dagger} \hat{b}_l^{\dagger} \rangle_B = 0$ and that $\langle \hat{b}_k \hat{b}_l^{\dagger} \rangle_B$ and $\langle \hat{b}_k^{\dagger} \hat{b}_l \rangle_B $ are non-zero if and only if $k = l$. 
If we further assume that the environment is thermal equilibrium at a temperature $T$, then $\hat{\rho}_B$ is represented by a thermal Gibbs state of the form
\begin{equation}
    \hat{\rho}_B = \frac{e^{- \beta \hat{H}_B }}{\mathcal{Z}_B} \, ,
\end{equation}
where $\mathcal{Z}_B$ is the reservoir partition function. As a result, we find that the quantity $\langle \hat{b}_k^{\dagger} \hat{b}_k \rangle_B = N_k = (e^{\beta \omega_k} - 1)^{-1}$ is the mean occupation number of the $k$-th mode of the environment. Finally, assuming that the bath modes form a continuum, we obtain the following expression for the correlation function:
\begin{equation}
     \alpha_{\beta} (t) = \int
     _{0}^{\infty} \textrm{d} \omega \,J(\omega) \left[ \coth \left( \frac{\beta \omega}{2} \right) \cos \left( \omega t \right){-}i \sin \left( \omega t \right) \right] \,
\end{equation}
which can be recast in the form of \Cref{eq:bosonic_corr_f1,eq:bosonic_corr_f2}.

\section{SB model (pure dephasing)}
\label{app:B}

We now derive the equations governing the dynamics of the system described in section \ref{subsec:exactlysolvable}. We work in the interaction picture and begin by deriving an expression for the unitary evolution operator $\hat{U} (t)$ which acts on the composite system.
Let us first notice that the two-time commutator of the interaction Hamiltonian is non-zero, i.e.
\begin{equation}
\label{eq:SB_commutator}
    \left[ \hat{H}_I (t) , \hat{H}_I (t') \right] = - 2 i \, \mathbb{1}_{S} \otimes \sum_k |g_k|^2 \sin (\omega_k (t - t')) ,
\end{equation}
where $\mathbb{1}_S$ is the identity operator acting on the system only. The latter is useful to evaluate the time evolution operators as
\begin{equation}
    \hat{U} (t) = \mathcal{T}_{\leftarrow} \exp\left[ -i \int_0^t \hat{H}_I (\tau) \,\textrm{d} \tau \right] \, ,
\end{equation}
where $\mathcal{T}_{\leftarrow}$ denotes the time ordering operator.
Following the ideas in Ref. \cite{Lidar:2001} (see also Ref.~\cite{Schaller:2014}), we can formally discretise the integral in the exponent of the unitary evolution operator and denote $\mathcal{H}_n = - i \hat{H}_I (n \textrm{d}t)$, where $\textrm{d}t = t/N$. Taking the limit as $N \to \infty$ we obtain
\begin{equation}
    \hat{U} (t) = \mathcal{T}_{\leftarrow} \lim_{\textrm{d}t \to 0} \exp\left[ \sum_{n = 1}^N \mathcal{H}_n\, \textrm{d}t\right] \, .
\end{equation}
We use a generalisation of the Baker-Campbell-Hausdorff formula to calculate the exponential
\begin{equation}
    e^{ \sum_{n=1}^N \mathcal{H}_n}  = \left( \prod_{n = 1}^N e^{ \mathcal{H}_n} \right) \left( \prod_{n < m } e^{ -  \frac{1}{2} \left[ \mathcal{H}_n, \mathcal{H}_m \right]} \right) \, ,
\end{equation}
which holds since the second order commutators vanish. The unitary evolution operator becomes
\begin{equation}
    \tilde{U} (t) = \lim_{\textrm{d}t \to 0 } \prod_{n < m } e^{  - \frac{1}{2} \left[  \mathcal{H}_n, \mathcal{H}_m \right] (\textrm{d}t)^2} \prod_n e^{ \mathcal{H}_n \textrm{d} t} \, ,
\end{equation}
where we have noticed that the commutator in the first exponent is just a complex number, so we may omit the time ordering operator. Recombining the exponentials of the operators we find
\begin{equation}
\begin{aligned}
    \tilde{U} (t) & = \lim_{\textrm{d}t \to 0} e^{ - \frac{1}{2} \sum_{n<m} \left[ \mathcal{H}_n, \mathcal{H}_m \right] (\textrm{d}t)^2} e^{ \sum_n \mathcal{H}_n \textrm{d} t }\\
    & = e^{ \frac{1}{2} \int_0^t \textrm{d}t_1 \int_0^{t_1} \textrm{d} t_2 \left[ \hat{H}_I (t_2), \hat{H}_I (t_1) \right]} e^{ -i \int_0^t \hat{H}_I (\tau) \textrm{d} \tau},
\end{aligned}
\end{equation}
where the first exponent --  as a consequence of \Cref{eq:SB_commutator} -- only applies a global phase to the qubit. As a result, the dynamics of the system are solely governed by the operator
\begin{equation}
    e^{-i \int_0^t \hat{H}_I (\tau) \textrm{d} \tau } = e^{\hat{\sigma}_z \otimes\sum_k \left( \alpha_k (t) \hat{b}_k^{\dagger} - \alpha_k^* (t) \hat{b}_k \right)} \equiv e^{\hat{\sigma}_z \otimes \hat{A}(t)} \, ,
\end{equation}
with $\alpha_k(t) = g_k \left( 1 - e^{i \omega_k t}\right)/\omega_k$. It is convenient to rewrite this operator in the form:
\begin{align}
    e^{\hat{\sigma}_z \otimes \hat{A}(t)} & = I \otimes \sum_{n=0}^{\infty} \frac{\hat{A}(t)^{2n}}{2n!} + \hat{\sigma}_z \otimes \sum_{n=1}^{\infty} \frac{\hat{A}(t)^{2n+1}}{\left( 2n+1 \right)!} \\
    & = I \otimes \cosh ( \hat{A}(t) ) + \hat{\sigma}_z \otimes \sinh (\hat{A} (t)) \, .
\end{align}
The matrix elements of the reduced density matrix are determined by explicitly tracing out the environmental degrees of freedom, i.e.
\begin{equation}
    \hat{\rho}_{ ij }(t) = \bra{i} \operatorname{tr}_B \left\{ \hat{U} (t) \hat{\rho}^0 \otimes \hat{\rho}_B \hat{U}^{\dagger} (t) \right\} \ket{j} \, .
\end{equation}
It follows that the coherences of the reduced density matrix evolve as 

\begin{equation}
    \hat{\rho}_{0 1}(t) = \hat{\rho}_{ 0 1}^0  \left \langle e^{2 \hat{A} (t)} \right \rangle \, ,
\end{equation} 
with $\hat{\rho}_{ 1 0 }(t) = \hat{\rho}_{ 0 1 }^*(t)$. Resorting to the identity $\langle e^{\hat{A}} \rangle = e^{\langle \hat{A} \rangle^2/2}$, where the operator $\hat{A}$ is a linear combination of creation and annihilation operators \cite{giuliani2005quantum}, we find that
\begin{align}
    \langle e^{ 2 \hat{A} (t)}\rangle & = e^{- 2 \sum_k | \alpha_k (t) |^2 \langle b_k b_k^{\dagger} + b_k^{\dagger} b_k \rangle} \nonumber \\
    & = e^{ - 2 \sum_k |\alpha_k (t)|^2 (2 N_k + 1 )} \, .
\end{align}
Finally, substituting the expressions for $\alpha_k$ and the mean occupation number of the $k$-th mode of the environment, $N_k$, we obtain
\begin{equation}
    \langle e^{ 2 \hat{A} (t)}\rangle = e^{- \Gamma (t)} \, ,
\end{equation}
where we have assumed the the bath modes form a continuum. The function $\Gamma (t)$ is the decoherence function given in \Cref{eq:decoherencefunction} of the main text.

\section{SB model (amplitude damping)}
\label{app:C}
Here, we derive the equations governing the dynamics of the system described in \Cref{subsec:nonMarkovSB}. The second-order generator of the TCL master equation leads to the following equation for the reduced density matrix in the interaction picture $\tilde{\hat{\rho}}$~\cite{Breuer-Petruccione1,Rivas2012}:
\begin{equation}
    \frac{d \tilde{\hat{\rho}}}{ d t} = - \int_0^t \textrm{d} s \operatorname{tr}_B \left[ \hat{H}_I (t), \left[ \hat{H}_I (s) , \tilde{\hat{\rho}} \otimes \hat{\rho}_B \right] \right] \, ,
\end{equation}
\\
where $\tilde{\hat{\rho}}$ and $\hat{H}_I (t) = - \hat{\sigma}_x (t) \otimes \hat{B}(t) /2 $ are expressed in the interaction picture with respect to the free Hamiltonian $H_S$. The form of the bath operator $\hat{B} (t)$ is given by  \Cref{eq:B_int_pic}. By explicitly performing the calculations, changing the integration variable as $s\to t - s$, and moving to the Schr\"odinger picture, we
are able to rewrite such master equation as

\begin{equation}
\begin{aligned}
    \frac{d \hat{\rho}}{d t} = & - i \left[ \hat{H}_S, \hat{\rho} \right] - \frac{1}{4} \int_0^t \textrm{d} s\,\left( \nu (s) \left[ \hat{\sigma}_x , \left[ \hat{\sigma}_x (-s) , \hat{\rho}  \right] \right]\right. \\
    &\left. + i \eta (s) \left[ \hat{\sigma}_x , \left\{ \hat{\sigma}_x (-s) , \hat{\rho} \right\} \right] \right),
\end{aligned}
\end{equation}

where $\nu (s)$ and $\eta (s)$ are respectively the real and imaginary parts of the correlation function given in \Cref{eq:bosonic_corr_f2} of the main text. The corresponding dynamical equations for the components of the Bloch vector $\langle {\sigma}_j (t)\rangle = \text{tr}_S\left[ {\sigma}_j \rho(t) \right]$ read

\begin{align}
    \frac{d \langle \hat{\sigma}_x (t) \rangle}{d t} &  = - \omega_0 \langle \hat{\sigma}_y (t) \rangle \, , \\
    \frac{d \langle \hat{\sigma}_y (t) \rangle}{d t} & =  \left( \omega_0 + a_{yx} (t) \right) \langle \hat{\sigma}_x (t) \rangle + a_{yy} (t) \langle \hat{\sigma}_y (t) \rangle \, ,  \\
    \frac{d \langle \hat{\sigma}_z (t) \rangle}{d t} & = a_{zz} (t) \langle \hat{\sigma}_z (t) \rangle + b_z (t) \, ,
\end{align}

where the time-dependent coefficients are defined in the main text [Cf. \Cref{eq:time_dep_coeff1,eq:time_dep_coeff2}]. This set of coupled differential equations can be recast in the matrix form of \Cref{eq:blochME,eq:A}.

\bibliographystyle{apsrev4-2.bst}
\bibliography{biblio.bib}

\end{document}